\documentstyle[preprint,aps,version2]{revtex}
\draft
\def\Journal#1#2#3#4{{#1} {\bf#2}, {#3} {(#4)}}
\def\PRL{Phys. Rev. Lett.}
\def\PRD{{Phys. Rev.} D}
\def\PLB{{Phys. Lett.}  B}
\def\EPJC{{Eur. Phys. J.} C}
\def\ZPC{{Z. Phys.} C}

\def\JPG{J. Phys. G}
\def\NP{{ Nucl. Phys.}}

\def\ra{\rightarrow}

\def\be{\begin{equation}}
\def\ee{\end{equation}}
\def\bea{\begin{eqnarray}}
\def\eea{\end{eqnarray}}

\def\qbar{{\bar q}}
\def\ubar{{\bar u}}
\def\dbar{{\bar d}}
\def\sbar{{\bar s}}
\def\etap{{\eta^\prime}}

\begin{document}
%{\Large

\begin{titlepage} 
%\begin{flushright}
%BIHEP-TH-96-30 \\
%November 1996
%\end{flushright}

\bigskip
\begin{title}
{\large\bf Two analytical constraints on the $\eta$-$\etap$ mixing}
\end{title}
\author{Fu-Guang Cao\thanks{E-mail address: f.g.cao@massey.ac.nz.}
and A. I. Signal\thanks{E-mail address: a.i.signal@massey.ac.nz.}
}
\begin{instit}
Institute of Fundamental Sciences, Massey University,
Private Bag 11 222, Palmerston North,
New Zealand
\end{instit}

\begin{abstract}
We obtained two analytical constraints on the $\eta$-$\etap$ mixing parameters
by considering two-photon decays of $\eta$ and $\etap$
[$\eta \, (\etap) \, \ra \gamma \gamma$],
and productions of $\eta$ and $\etap$ in the $e^+ e^-$ scattering
at large momentum transfer ($Q^2 \ra \infty$).
Using the data given in the PDG98 for the decay processes and
recent CLEO measurements on the meson-photon transition form factors,
we estimate for the $\eta_8$-$\eta_1$ mixing scheme
the mixing angle to be $\theta=-14.5^\circ \pm 2.0^\circ$
and the ratio of the decay constants of singlet to octet
to be $f_1/f_8=1.17 \pm 0.08$.
Applying our approach to the recently proposed
$q \qbar$-$s \sbar$ mixing scheme,
we obtain the mixing angle to be $\phi=39.8^\circ \pm 1.8^\circ$
and the ratio of the decay constants of $s\sbar$ state to $q\qbar$ state
to be $f_s/f_q=1.20 \pm 0.10$.

\bigskip
\noindent
PACS number(s): 14.40.Aq; 11.40.Ha; 13.40.Gp; 12.38.Bx
\end{abstract}
\end{titlepage}

As is well known, the $SU(3)$ quark model predicts the existence of an octet of
massless pseudoscalar $\eta_8=\frac{1}{\sqrt{6}}(u\ubar+d\dbar-2\, s\sbar)$
and a massive singlet $\eta_1=\frac{1}{\sqrt{3}}(u\ubar+d\dbar+s\sbar)$.
The physical states $\eta$ and $\etap$ arise from the mixing between
$\eta_8$ and $\eta_1$,
\bea
\left(
	\begin{array}{c}
	\eta \\
	\etap
	\end{array}
\right)
=\left(
	\begin{array}{cc}
	{\rm cos} \, \theta & -{\rm sin} \, \theta \\
	{\rm sin} \, \theta & {\rm cos} \, \theta
	\end{array}
\right)
\left(
	\begin{array}{c}
	\eta_8 \\
	\eta_1
	\end{array}
\right).
\label{Mixing1}
\eea
Studying the $\eta$-$\etap$ mixing is important to the understanding of
quark model and QCD. There have been many studies on the mixing angle
$\theta$, see e.g.
\cite{Gasser,Donoghue,Gilman,Schechter,Kisselev,Ball,Bramon,Kaiser,Burakovsky,Jakob,Choi,Feldmann0,Feldmann1,JCao}
and references therein.
The recent experimental measurements on the productions of $\eta$ and
$\etap$ from the CLEO collaboration \cite{CLEO}
and L3 collaboration \cite{L3} have stimulated more
phenomenological studies on this issue \cite{Choi,Feldmann0,Feldmann1,JCao}.
The theoretical analyses are usually model dependent,
and the predictions for $\theta$ vary from $-12^\circ$ \cite{Schechter,Feldmann1}
to $-20^\circ$ \cite{Donoghue,Burakovsky}.
Besides the mixing angle $\theta$, another two parameters $f_8$ and $f_1$,
the decay constants of $\eta_8$ and $\eta_1$, are usually introduced.
Model analyses estimate $f_8$ in the range
of $0.71 f_\pi$ \cite{Kisselev}\footnote{We quoted the value
evaluated from \cite{Kisselev} by the authors of Ref.~\cite{Feldmann1}.}
to $1.28 f_\pi$ \cite{Donoghue,Kaiser,Feldmann0,Feldmann1}
and $f_1$ in the range of $0.94 f_\pi$ \cite{Kisselev}
to $1.25 f_\pi$ \cite{Kaiser}.
Once again, the uncertainty in the model predictions is sizable.

Recently a two-mixing-angle scheme
\cite{Schechter,Kisselev,Kaiser,Feldmann0}
was investigated, especially in the studies of the mixing of decay constants,
\bea
\left(
	\begin{array}{c}
	f_\eta \\
	f_\etap
	\end{array}
\right)
=\left(
	\begin{array}{cc}
	{\rm cos} \, \theta_8 & -{\rm sin} \, \theta_1 \\
	{\rm sin} \, \theta_8 & {\rm cos} \, \theta_1
	\end{array}
\right)
\left(
	\begin{array}{c}
	f_{\eta_8} \\
	f_{\eta_1}
	\end{array}
\right).
\label{Mixing2}
\eea
Although it was frequently
assumed \cite{Donoghue,Gilman,Ball,Bramon,Burakovsky,JCao}
that the decay constants follow the pattern of state mixing [Eq.~(\ref{Mixing1})],
the mixing properties of the decay constants will be generally different from
the mixing properties of the meson state since the decay constants are
controlled by the specific Fock state wave functions at zero spatial separation
of the quarks while the state mixing refers to the mixing of the overall
wave functions \cite{Feldmann1}.
The two-angle-mixing scheme, Eq.~(\ref{Mixing2}), is suitable
in the studies of the mixing of decay constants, while 
the common $\eta_8$-$\eta_1$ mixing scheme, Eq.~(\ref{Mixing1}),
is applicable in the studies of state mixing 
since only one mixing angle is required in this situation.
More recently a mixing scheme based on the quark flavour basis
$q \qbar=(u \ubar + d \dbar)/\sqrt{2}$ and $s\sbar$ was proposed \cite{Feldmann1}.
It was assumed that the decay constants follow the state mixing if and only if they
are defined with respect to the $q \qbar$-$s \sbar$ basis,
\bea
\left(
	\begin{array}{c}
	\eta (f_\eta)\\
	\etap (f_\etap)
	\end{array}
\right)
=\left(
	\begin{array}{cc}
	{\rm cos} \, \phi & -{\rm sin} \,\phi \\
	{\rm sin} \, \phi & {\rm cos} \,\phi
	\end{array}
\right)
\left(
	\begin{array}{c}
	\eta_q (f_{\eta_q})\\
	\eta_s (f_{\eta_s})
	\end{array}
\right),
\label{Mixing3}
\eea
where $\phi$ is the mixing angle.
One advantage of the $q\qbar$-$s\sbar$ mixing scheme is that
only one mixing angle is introduced but not two as that in Eq.~(\ref{Mixing2}).

In this letter,
we concern ourselves about the mixing of particle states
[the mixing of decay amplitudes (see Eqs.~(\ref{w1}) and (\ref{w2}))
and the mixing of meson-photon transition form factors
(see Eqs.~(\ref{Feta}) and (\ref{Fetap}))]
but not the mixing of decay constants.
Thus we employ the $\eta_8$-$\eta_1$ mixing scheme [Eq.~(\ref{Mixing1})]
and the $q\qbar$-$s\sbar$ mixing scheme [Eq.~(\ref{Mixing3})].
We emphasize that in our analyses
the decay constants are not implied to follow the same mixing schemes
as the particle states, {\it i.~e.} Eqs.~(\ref{Mixing1}) and (\ref{Mixing3})
will not be applied to the mixing of the decay constants.
It has been well known \cite{Jakob,Feldmann0,JCao}
that the two-photon decays of $\eta$ and $\etap$ and the
production processes of $\eta$ and $\etap$ in the $e^+e^-$ scattering
can be used to extract the mixing parameters.
However, we will analyses these exclusive processes in
a different way from that in Refs.~\cite{Jakob,Feldmann0,JCao}:
Considering the ratio of the two-photon decay widths of $\eta$ and $\etap$
and the ratio of the $\eta$-$\gamma$ and $\etap$-$\gamma$ transition factors
at the large momentum transfer limit ($Q^2 \ra \infty$), we find that
the above ratios satisfy two analytical equations which involved
the same two mixing parameters --
the mixing angle and the ratio of decay constants.
From the two equations we obtain analytical expressions for the mixing angle
($\theta$ or $\phi$)
and the ratio of the decay constants ($f_1/f_8$ or $f_s/f_q$) for
the $\eta_8$-$\eta_1$ mixing and $q\qbar$-$s\sbar$ mixing schemes respectively.

%The constraints on the recently proposed $q \qbar$-$s \sbar$
%mixing scheme \cite{Feldmann1} are also discussed.
%Applying our approach to the recently proposed $q \qbar$-$s \sbar$
%mixing scheme \cite{Feldmann1} results in two constraints on
%the corresponding mixing angle $\phi$ and the ratio of the decay constants
%of the $s\sbar$ state to $q\qbar$ state, $f_s/f_q$.

The first constraint comes from the two-photon decays of $\eta$ and $\etap$.
The decay amplitudes of $\pi^\circ \ra \gamma \gamma$,
$\eta_8 \ra \gamma \gamma$ and $\eta_1 \ra \gamma \gamma$
have the same Lorentz structure \cite{DonoghueBook},
\bea
A_{P\ra r(k_1) r(k_2)}=\frac{\alpha}{\pi} \,\frac{c_P}{f_P}
\, \epsilon^{\mu \nu \alpha \beta} \epsilon^*_\mu(k_1) \epsilon^*_\nu(k_2)
k_{1\alpha} k_{2 \beta},
\label{A}
\eea
where
\bea
c_P=\left( \, 1, \, \frac{1}{\sqrt{3}}, \, \frac{2\sqrt{2}}{\sqrt{3}} \,\right)
\label{cM}
\eea
for the unmixed states $P=(\pi^0,\, \eta_8,\, \eta_1)$
and $f_P$ is the corresponding decay constant ($f_\pi=93$ MeV).
Generalizing the PCAC result for $\pi^0 \ra \gamma \gamma$ to the decays
$\eta_8 \ra \gamma \gamma$ and $\eta_1 \ra \gamma \gamma$,
and assuming the mixing occurs at amplitude level \cite{Donoghue},
we can obtain
\bea
\Gamma_{\pi^0 \ra \gamma \gamma}
&=&\frac{\alpha^2 \,m_{\pi^0}^3}{64 \,\pi^3} 
\left[\frac{c_{\pi^0}}{f_\pi}\right]^2,
\label{w0} \\
\Gamma_{\eta \ra \gamma \gamma}
&=&\frac{\alpha^2\,m_\eta^3}{64 \,\pi^3} 
\left[\frac{c_8 \,{\rm cos} \, \theta}{f_8}
-\frac{c_1 \,{\rm sin} \, \theta}{f_1} \right]^2,
\label{w1}\\
\Gamma_{\etap \ra \gamma \gamma}
&=&\frac{\alpha^2\,m_{\etap}^3}{64 \,\pi^3} 
\left[\frac{c_8 \,{\rm sin} \, \theta}{f_8}
+\frac{c_1 \,{\rm cos} \, \theta}{f_1} \right]^2.
\label{w2}
\eea
Combining Eqs. (\ref{w1}) and (\ref{w2}) yields
\bea
\frac{\Gamma_{\eta \ra \gamma \gamma}}
{\Gamma_{{\etap} \ra \gamma \gamma}}
=\frac{m_\eta^3}{m_{\etap}^3}
\left[\frac{f_1/f_8 -c_1/c_8 \,{\rm tan} \, \theta}
{f_1/f_8 \,{\rm tan} \, \theta +c_1/c_8}
\right]^2.
\label{w1over2}
\eea
It is worth noting that only two parameters, $\theta$ and $f_1/f_8$,
appear in Eq.~(\ref{w1over2}) while three parameters, $\theta$,
$f_1$ and $f_8$, are involved in Eqs.~(\ref{w1}) and (\ref{w2}).
If we now let
\bea
c&=&\frac{c_1}{c_8} =\sqrt{8},
\label{c}\\
r&=& \frac{f_1}{f_8},
\label{r}\\
\rho_1&=&\left[
\frac{\Gamma_{\eta \ra \gamma \gamma}}
{\Gamma_{{\etap} \ra \gamma \gamma}}
\,\frac{m_{\etap}^3}{m_\eta^3}
\right]^{1/2},
\label{rho1}
\eea
we obtain the first constraint on the parameters $\theta$ and $r$
from Eq.~(\ref{w1over2})
\bea
{\rm tan} \, \theta =\frac{r - c \,\rho_1}{c +r \,\rho_1}.
\label{constraint1}
\eea

The second constraint comes from the productions of $\eta$ and ${\etap}$
in the $e^+ e^-$ scattering, $e^+ e^- \ra e^+ e^-  \eta ({\etap})$
$[\gamma \gamma^* \ra \eta (\etap)]$, at large momentum transfer.
This class of process can be described using only one form factor,
the meson-photon transition form factor $F_{P \gamma}(Q^2)$.
It has been noted \cite{Jakob,Feldmann0,JCao} that this form factor
may provide useful information on the $\eta$-$\etap$ mixing.
The common procedure to extract the $\eta$-$\etap$ mixing angle from
these processes is to fit the perturbative calculations for the transition
form factors to the experimental data in the range of $Q^2$ being larger than,
say, $1$ or $2$ GeV$^2$.
At present the available experimental data for the $F_{\eta \gamma}(Q^2)$
and $F_{\etap \gamma}(Q^2)$ are in the ranges of $Q^2 < 20$~GeV$^2$
and $Q^2 < 30$~GeV$^2$, respectively \cite{CLEO,L3}.
As we know, although perturbative theory can make reliable
prediction for the asymptotic ($Q^2 \ra \infty$) behavior of exclusive
processes, the perturbative calculations in the currently experimentally
accessible energy region, especially in the lower energy end of the 
experimental data, may suffer from large corrections such as
higher order contributions in $\alpha_s(Q^2)$
and higher twist effects.
Also the perturbative calculations usually employ some model wave functions
to account for the non-perturbative properties
of $\eta$ and $\etap$ \cite{Jakob,Feldmann0,JCao}.
Thus the usual procedure to extract the $\eta$-$\etap$ mixing angle
from these production processes has large uncertainty, and
even the reliability of this procedure was questioned
in Ref.~\cite{JCao} by studying these form factors in light-cone perturbative
theory. Here we would like to adopt an alternative scheme:
combining the well established perturbative predictions for the $\eta$ and
$\etap$ transition form factors in the asymptotic limit ($Q^2 \ra \infty$)
with the phenomenological formulas given by the CLEO collaboration
and L3 collaboration to obtain another analytical constraint on
the mixing angle $\theta$ and the ratio $f_1/f_8$.

The $Q^2 \ra \infty$ behavior of $F_{P \gamma}(Q^2)$
is well predicted by perturbative QCD \cite{Lepage,FGCao}.
For the $\pi^0$, $\eta_8$, and $\eta_1$, we have
\bea
F_{P \gamma}(Q^2)=\frac{2}{\sqrt{3}} \,c_P \int_0^1 dx \,
\frac{\phi_P (x)}{x (1-x) Q^2},
\label{FQ}
\eea
where $x$ and $1-x$ are the longitudinal momentum fractions carried by
the quark and
anti-quark in the meson respectively, and $c_P$ is given by Eq.~(\ref{cM}).
In the $Q^2\ra \infty$ limit, any meson distribution amplitudes approach
the asymptotic form
\bea
\phi_P(x)=\sqrt{3} \, f_P \, x (1-x).
\label{phi}
\eea
Thus we have
\bea
F_{P\gamma}(Q^2 \ra \infty)=\frac{2 c_P f_P}{Q^2}.
\label{FInfty}
\eea
Assuming the mixing occurs at the states, we have
\bea
F_{\eta \gamma}(Q^2 \ra \infty) &=&\frac{2 c_8 f_8}{Q^2} \,{\rm cos}\,\theta
					  		     -\frac{2 c_1 f_1}{Q^2} \,{\rm sin}\,\theta ,
\label{Feta} \\
F_{\etap \gamma}(Q^2 \ra \infty)&=&\frac{2 c_8 f_8}{Q^2} \,{\rm sin}\,\theta
					    		      +\frac{2 c_1 f_1}{Q^2} \,{\rm cos}\,\theta.
\label{Fetap}
\eea
If we let
\bea
\rho_2=\frac{F_{\eta \gamma}(Q^2 \ra \infty)}
{F_{\etap \gamma}(Q^2 \ra \infty)},
\label{rho2}
\eea
we can obtain the second constraint on the mixing angle $\theta$
and the ratio $f_1/f_8$ from Eqs.~(\ref{Feta}) and (\ref{Fetap}),
\bea
{\rm tan} \, \theta=\frac{1 - c \,r \rho_2}{c \,r + \rho_2}.
\label{constraint2}
\eea
We would like to emphasize that Eq.~(\ref{constraint2}) is an analytical
expression which depends on the experimental information on the ratio
of transition form factors $F_{\eta \gamma}(Q^2)/F_{\etap \gamma}(Q^2)$
at large momentum transfer.
Eq.~(\ref{constraint2}) suffers from experimental uncertainties on the
measurements of $F_{\eta \gamma}(Q^2)$ and $F_{\etap \gamma}(Q^2)$
at large (but finite) momentum transfer and
the systematic uncertainties associated with the
extrapolation of $F_{\eta \gamma}(Q^2)$ and $F_{\etap \gamma}(Q^2)$
from experimentally accessible energy region to infinity [see Eq.~(\ref{FQexp})].
The usual procedure to extract the mixing angle from these form factors
suffer from large corrections to the perturbative calculations in
the energy region of a few GeV$^2$ as well as the experimental uncertainty.

From Eqs.~(\ref{constraint1}) and (\ref{constraint2}), we obtain the following
analytical expressions for the mixing $\theta$ and ratio $f_1/f_8$,
\bea
{\rm tan}\, \theta =\frac{-(1+c^2)(\rho_1+\rho_2)
	+\sqrt{(1+c^2)^2(\rho_1+\rho_2)^2
	+4 (c^2-\rho_1 \rho_2)(1-c^2\rho_1 \rho_2)}}
	{2(c^2 -\rho_1 \rho_2)},
\label{tantheta}
\eea
\bea
r = \frac{(1+c^2)(\rho_1 -\rho_2)
	+\sqrt{(1+c^2)^2(\rho_1 - \rho_2)^2+4 c^2(1+\rho_1 \rho_2))^2}}
	{2 c(1 +\rho_1 \rho_2)}.
\label{rresult}
\eea

From Eqs.~(\ref{w0}) and (\ref{w1}), we can also evaluate the ratios
$f_8/f_\pi$ and $f_1/f_\pi$,
\bea
\frac{f_8}{f_\pi}&=&\rho_0
\left[
\frac{c_8}{c_\pi} \, {\rm cos} \,\theta
 - \frac{1}{r}\,\frac{c_1}{c_\pi}\, {\rm sin}\, \theta
\right],
\label{f8overpi}\\
\frac{f_1}{f_\pi}&=&\rho_0
\left[
\frac{c_8}{c_\pi} \,r \, {\rm cos} \, \theta 
- \frac{c_1}{c_\pi}\, {\rm sin}\, \theta
\right],
\label{f1overpi}
\eea
where
\bea
\rho_0=\left[
\frac{\Gamma_{\pi^0 \ra \gamma \gamma}}
{\Gamma_{{\eta} \ra \gamma \gamma}}
\,\frac{m_\eta^3}{m_{\pi^0}^3} \right]^{1/2}.
\label{rho0}
\eea

The parameters $\rho_0$ and $\rho_1$ [see Eq.~(\ref{rho1})] can be fixed
by using the two photon decay widths of $\pi^0$, $\eta$ and $\etap$ and
their masses. We employ the data given in the PDG98 \cite{PDG98},
\bea
\Gamma_{\pi^0 \ra \gamma \gamma}= 7.74 \pm 0.55 {\rm eV}, \,\,\,\,\,
&m_{\pi^0} =134.9764 \pm 0.0006 {\rm MeV},
\label{datapi0}\\
\Gamma_{\eta \ra \gamma \gamma}= 0.46 \pm 0.04 {\rm keV}, \,\,\,
&m_\eta =547.30 \pm 0.12 {\rm MeV},
\label{dataeta}\\
\Gamma_{\etap \ra \gamma \gamma}= 4.37 \pm 0.25 {\rm keV}, \,\,\,
&m_{\etap} =957.78 \pm 0.14 {\rm MeV}.
\label{dataetap}
\eea
The parameter $\rho_2$ [see Eq.~(\ref{rho2})]
can be determined by using the experimental data
on the transition form factors at large momentum transfer.
Recently, the CLEO collaboration \cite{CLEO} has measured
the $F_{\eta \gamma}(Q^2)$ and $F_{\etap \gamma}(Q^2)$
in the $Q^2$ regions up to $20$ and $30$ GeV$^2$ respectively,
and the L3 collaboration \cite{L3} has measured
the $F_{\etap \gamma}(Q^2)$ in the $Q^2$ range up to 10 GeV$^2$.
Both CLEO and L3 present their results for the
transition form factor by a pole form \cite{CLEO,L3}
\bea
F_{P \gamma}(Q^2) = \frac{1}{4 \pi \alpha}\left[
	\frac{ 64 \pi \Gamma_{P \ra \gamma \gamma}}{m_P^3} \right]^{1/2}
	\frac{1}{1+Q^2/\Lambda_P^2},
\label{FQexp}
\eea
where $\Lambda_P$ is the pole mass parameter.
Thus we have
\bea
\rho_2=\left. \left[ \frac{\Gamma_{\eta\ra\gamma\gamma}}
	{\Gamma_{\etap\ra\gamma\gamma}}
	\,\frac{m_{\etap}^3}{m_\eta^3} \right]^{1/2}
	\frac{1 + Q^2/\Lambda_{\etap}^2}
	{1+Q^2/\Lambda_\eta^2} \right |_{Q^2 \ra \infty}
 =\rho_1 \frac{\Lambda_\eta^2}{\Lambda_{\etap}^2}.
\eea
The CLEO results for the pole-mass parameters \cite{CLEO}
are\footnote{We will not use the result of L3 collabration
since only one pole-mass parameter, $\Lambda_\etap$,
was given in \cite{L3}.}
\bea
\Lambda_\eta=774 \pm 11 \pm 16 \pm 22 \,{\rm MeV}, \,\,\,\,
\Lambda_{\etap}=859 \pm 9 \pm 18 \pm 20 \,{\rm MeV},
\label{Lambdadata}
\eea
where the first error represents statistical, the second error is systematic,
and the third error comes from the uncertainty in the value of
$\Gamma_{\eta \, (\etap) \ra \gamma \gamma}$.
Using the data presented in Eqs.~(\ref{datapi0}), (\ref{dataeta}),
(\ref{dataetap}) and (\ref{Lambdadata}), we obtain
\bea
\theta= -14.5^\circ \pm 2.0^\circ, \,\,\,\,
&f_1/f_8= 1.17 \pm 0.08,
\label{rtheta} \\
f_8/f_\pi= 0.94 \pm 0.07, \,\,\,\,
&f_1/f_\pi= 1.09 \pm 0.05,
\label{fratios1}
\eea
where we have added all errors in quadrature in our calculations.

Our prediction for the mixing angle $\theta= -14.5^\circ \pm 2.0^\circ$
is comparable with the result $\theta\simeq -17^\circ$ given
in Refs.~\cite{Ball,Bramon} which are obtained
by considering various decay processes, especially the $J/\Psi$ decays.
Also our prediction is larger than the prediction
$\theta\simeq -23^\circ$ \cite{Donoghue,Burakovsky}
based on the Chiral Lagrangian and phenomenological mass formulas
(Gell-Mann-Okubo formula etc.).
We notice that the mixing of decay constants was assumed
to follow the same pattern of the state mixing
in \cite{Donoghue,Burakovsky}
which is questionable as we mentioned in the introduction.
Thus the inconsistency between our prediction for the mixing
angle and the prediction from Chiral theory
\cite{Donoghue,Burakovsky} is not as serious as it seems to be.
We will not compare our prediction with the results from
the two-mixing-angle schemes \cite{Schechter,Kisselev,Kaiser,Feldmann0}
since different mixing schemes are employed.
Our result for the ratio $f_8/f_\pi$ being $0.94 \pm 0.07$ is smaller than
the prediction from Chiral perturbation theory (ChPT)
$f_8/f_\pi=1.28$ \cite{Donoghue,Kaiser}
and most phenomenological analyses
$f_8/f_\pi=1.2 - 1.3$ \cite{Gasser,Gilman,Burakovsky,Feldmann0,Feldmann1},
but is larger than the result given in Ref.~\cite{Kisselev} $f_8/f_\pi=0.71$.
Our prediction for the ratio $f_1/f_\pi$ being $1.09 \pm 0.05$
is consistent with most phenomenological analyses being about $1.15$
\cite{Gasser,Gilman,Burakovsky,Feldmann0,Feldmann1}
and the ChPT prediction $f_1/f_\pi=1.05\pm 0.04$ given in \cite{Donoghue}
but is large than the ChPT prediction $f_1/f_\pi=1.25$ given
in \cite{Kaiser}.
In the previous studies either the questionable assumption
that the decay constants and the particle states share the same mixing scheme
\cite{Gasser,Donoghue,Gilman,Burakovsky} or two mixing-angle scheme
\cite{Kisselev,Kaiser,Feldmann0,Feldmann1} is adopted.
The relations between the mixing parameters involved in the two-mixing-angle
scheme and that appear in our model are remained to be further studied.
Thus our comparisons for the decay constants here are just suggestive.

Now we turn to the constraints on the parameters involved in the
$q \qbar$-$s \sbar$ mixing scheme [see Eq.~(\ref{Mixing3})].
The center assumption in this mixing scheme is
that the decay constants follow the state mixing if and only if they
are defined with respect to the $q \qbar$-$s \sbar$ basis \cite{Feldmann1}.
We would like to point out that our analysis for the $\eta_8$-$\eta_1$ mixing
scheme can be easily applied to the $q\qbar$-$s\sbar$ mixing scheme
by replacing the parameters $c=c_1/c_8$ and $r=f_1/f_8$
in Eqs.~(\ref{tantheta}) and (\ref{rresult})
with $c^\prime=c_s/c_q=\sqrt{2}/5$ and $r^\prime=f_s/f_q$.
Employing the data for the two-photon decay of $\pi^0$, $\eta$ and $\etap$
given in the PDG98 [see Eqs.~(\ref{datapi0}), (\ref{dataeta}) and (\ref{dataetap})]
and the CLEO result for the transition form factors [see Eq.~(\ref{Lambdadata})],
we can obtain
\bea
\phi=39.8^\circ \pm 1.8^\circ, \,\,\,\,
&f_s/f_q= 1.20 \pm 0.10,
\label{rphi} \\
f_q/f_\pi=1.06 \pm 0.05, \,\,\,\,
&f_s/f_\pi= 1.27 \pm 0.12.
\label{fratios2}
\eea
Our predictions [Eqs.~(\ref{rphi}) and (\ref{fratios2})] are consistent with
the phenomenological results $\phi=39.3^\circ \pm 1.0^\circ$,
$f_q/f_\pi=1.07 \pm 0.02$ and $f_s/f_\pi=1.34 \pm 0.06$
given in Ref.~\cite{Feldmann1}, which implies that as one concerns about
the state mixing the one-mixing angle schemes,
the $\eta_8$-$\eta_1$ mixing [Eq.~(\ref{Mixing1})]
and the $q\qbar$-$s\sbar$ mixing [Eq.~(\ref{Mixing3})], are applicable.
This consistency also implies that our predictions for the $\eta_8$-$\eta_1$
mixing parameters [Eqs.~(\ref{rtheta}) and (\ref{fratios1})] are reliable.

In summary,
we obtained two analytical constraints on the $\eta$-$\etap$ mixing parameters
by considering the two-photon decays
$\eta \, (\etap) \, \ra \gamma \gamma$
and the productions of $\eta$ and $\etap$ in the $e^+ e^-$ scattering
at large momentum transfer ($Q^2 \ra \infty$).
One advantage of our analysis is that it can be easily updated with any new
experimental data on the decay widths and meson-photon transition
form factors.
Using the data given in the PDG98 for the decay processes and
recent CLEO measurements on the meson-photon transition form factors,
we obtain $\theta=-14.5^\circ \pm 2.0^\circ$, $f_1/f_8=1.17 \pm 0.08$
for the $\eta_8$-$\eta_1$ mixing scheme,
and $\phi=39.8^\circ \pm 1.8^\circ$,  $f_s/f_q=1.20 \pm 0.10$
for the recently proposed $q \qbar$-$s \sbar$ mixing scheme.

%\end{document}

\section*{Acknowledgments}
This work was partially supported by the Massey Postdoctoral Foundation,
New Zealand.

%\newpage

%}
\end{document}